\documentclass{ctr_summer}

\usepackage{ctrfont}

\usepackage[dvips]{graphicx}
\usepackage{psfrag}
\usepackage{amsmath,rotating}

\newcommand{\bec}[1]{\mbox{\boldmath $ #1$}}

\title{LES of turbulent convection in solar-type stars and formation of large-scale magnetic structures}
\shorttitle{LES of turbulent convection in solar-type stars}

\author{I.~Rogachevskii, N.~Kleeorin \footnote{Department of Mechanical Engineering, Ben-Gurion University of the Negev} I.~N.~Kitiashvili, A.~G.~Kosovichev \footnote{Hansen Experimental Physics Laboratory, Stanford University}
A.~A.~Wray \and N.~N.~Mansour \footnote{NASA Ames Research Center}}

\shortauthor{Rogachevskii et al.}

\begin{document}

\setcounter{page}{1}

\maketitle

One of the fundamental problems of solar magnetism is understanding
the formation of long-lived compact magnetic structures, such as
sunspots and pores, in the highly turbulent upper layer of the solar
convective zone. Observations show that the process of accumulation
of fragmented magnetic fields emerging from the interior into these
compact structures occurs in the upper layers of the convection zone.
In this study we investigate the effects of turbulent convection on formation of large-scale inhomogeneous magnetic structures by means of Large-Eddy Simulation (LES) for convection in solar-type stars. The main idea of this study is the implementation of a new subgrid-scale model for the effective Lorentz force in a three-dimensional nonlinear radiative magnetohydrodynamics (MHD) code developed for simulating the upper solar convection zone and lower atmosphere. To this end we derived the energy budget equations, which include the effects of the subgrid-scale turbulence on the Lorentz-force, and implemented the new subgrid-scale turbulence model (TELF-Model) in a three-dimensional nonlinear MHD LES code. Using imposed initial vertical and horizontal uniform magnetic fields in LES with the TELF-Model, we have shown that the magnetic flux tubes formation is started when the initial mean magnetic field is larger than a threshold value (about 100 G). This is in  agreement with the theoretical studies by Rogachevskii \& Kleeorin (2007). We have determined the vertical profiles of the velocity and magnetic fluctuations, total MHD energy and anisotropy of turbulent magneto-convection, kinetic and current and cross helicities.
\vskip0.1in
\hrule

\section{Introduction}
\label{s:intro}

Significant progress toward quantitative understanding of the origin and nature of solar activity requires large-scale, integrated modeling of the physical conditions in subsurface layers of the Sun, where the magnetic field is generated and organized. Realistic MHD numerical simulations of subsurface flows and magnetic structures have become achievable because of the development of fast supercomputer systems and efficient parallel computer codes. In conjunction with modern subgrid-scale turbulence models, these numerical simulations provide an important tool for investigating the complicated physics of the upper turbulent convective boundary layer of the Sun. The dynamics of this layer is particularly critical for understanding the mechanism of formation of solar magnetic structures and predicting solar activity. Sunspots and pores represent one of the oldest and most intriguing problem of solar magnetism. Despite the long history of observational and theoretical investigations, the mechanism of their formation is still an open question.

The goal of this study is to investigate the effects of turbulent convection on the formation of large-scale inhomogeneous magnetic structures by means of Large-Eddy Simulation (LES) for convection in solar-type stars. The key idea of this study is the implementation of a new subgrid-scale model for effective Lorentz force in a three-dimensional, nonlinear radiative MHD code developed for simulating the upper solar convection zone and lower atmosphere (Jacoutot {\it et al.} 2008a,b; Kitiashvili {\it et al.} 2009, 2010). This new subgrid-scale model for the effective Lorentz force is based on

\noindent \textbullet~theoretical findings of the effects of turbulence
on the mean Lorentz force (Kleeorin \& Rogachevskii 1994; Kleeorin {\it et al.} 1996; Rogachevskii \& Kleeorin 2007),

\noindent \textbullet~direct numerical simulations (DNS) of this effect for (i) forced stratified turbulence (Brandenburg {\it et al.} 2010a,b) and (ii) turbulent convection (K\"{a}pyl\"{a} {\it et al.} 2010),

\noindent \textbullet~mean-field numerical modelling  (Brandenburg {\it et al.} 2010a,b; K\"{a}pyl\"{a} {\it et al.} 2010) based on parameterizations both of analytic formulae by Rogachevskii \& Kleeorin (2007) and the results of the DNS.

The physics of the effects of turbulence on the mean Lorentz force is as follows. The combined effects of the turbulent Reynolds and Maxwell stress tensors can lead to a local reduction of the total turbulent pressure and hence to the possibility of self-induced concentrations of large-scale magnetic fields (Rogachevskii \& Kleeorin 2007). Such a process may play an important role in the formation of sunspots and active regions in the Sun. DNS of forced stratified turbulence and turbulent convection have demonstrated that an imposed mean magnetic field results in a decrease of the effective magnetic pressure (Brandenburg {\it et al.} 2010a,b; K\"{a}pyl\"{a} {\it et al.} 2010). This phenomenon is quantified by determining the relevant contributions that relate the sum of the turbulent Reynolds and Maxwell stresses with the Maxwell stress of the mean magnetic field.

Using such a parameterization, Brandenburg {\it et al.} (2010a,b) and K\"{a}pyl\"{a} {\it et al.} (2010) have shown by means of the two-dimensional and three-dimensional mean-field numerical modelling that an isentropic (or isothermal) density stratified layer becomes unstable in the presence of a uniform imposed magnetic field. This large-scale instability results in the formation of loop-like magnetic structures that are concentrated at the top of the stratified layer. In three dimensions these structures resemble the appearance of bipolar magnetic regions in the Sun. The results of DNS and mean-field numerical modelling (Brandenburg {\it et al.} 2010a,b; K\"{a}pyl\"{a} {\it et al.} 2010) are in good agreement with the theoretical predictions by Rogachevskii \& Kleeorin (2007).

In this study we investigate formation of inhomogeneous magnetic structures using a three-dimensional radiative
MHD code, "SolarBox,'' developed by A. Wray at NASA Advanced
Supercomputing Division (Jacoutot {\it et al.} 2008a,b; Kitiashvili {\it et al.} 2009, 2010). This code takes into account several physical phenomena: compressible fluid flow in a highly stratified medium; three-dimensional multigroup radiative energy transfer between the fluid elements; a real-gas equation of state, ionization and excitation of all abundant species; and magnetic effects. The code is built for simulations of fluid flows in top layers of the convective zone
and the low atmosphere, in the rectangular geometry. An important feature of this code is implementation of various subgrid-scale turbulence models, e.g., the most widely used Smagorinsky model (Smagorinsky 1963) in the compressible formulation (Moin {\it et al.} 1991; Germano {\it et al.} 1991). The turbulent electrical conductivity in the code is determined by using the extension of the Smagorinsky model to the MHD case (Theobald {\it et al.} 1994). The code is based on the LES approach, and solves the grid-cell-averaged equations of the conservation of mass, momentum, and energy. We modify equations of this LES code by implementing a new subgrid-scale model for effective Lorentz force, and study the dynamics of magnetic structure formation from initial horizontal and vertical uniform magnetic fields in a turbulent convection.

Most of numerical simulations on magnetic flux emerging (see, e.g., Rempel {\it et al.} 2009, Stein {\it et al.} 2010) have been done using initial conditions with already existing strongly inhomogeneous
large-scale magnetic field. On the other hand, formations of the large-scale inhomogeneous magnetic structures from initial {\em uniform} horizontal or vertical magnetic fields in a small-scale turbulent convection have been studied in LES by Kitiashvili {\it et al.} (2010) and in DNS by Brandenburg {\it et al.} (2010a), Brandenburg {\it et al.} (2010b) and K\"{a}pyl\"{a} {\it et al.} (2010).

\section{Derivation of budget equation for energy of mean fields in small-scale MHD turbulence}

The effect of the MHD turbulence on the mean (large-scale) magnetic force is described by the turbulent magnetic coefficients $Q_p=1-q_p$ and $Q_s = 1-q_s$, so that the mean Lorentz force reads
\begin{eqnarray}
{\bf F}_m^{eff} = -\bec{\nabla}\biggl( {Q_p \over 8\pi}  {\bf
B}^2 \biggr) +  ({\bf B} \cdot \bec{\nabla}) {Q_s \over 4\pi} {\bf
B} ,
\label{C1}
\end{eqnarray}
(see Rogachevskii \& Kleeorin 2007), where ${\bf B}$ is the mean magnetic field, and the functions $q_p$ and $q_s$ determine the effect of small-scale MHD turbulence on the mean Lorentz force. Thus, the equations for the large-scale fields have the
following form:
\begin{eqnarray}
\rho \biggl({\partial \over \partial t} &+& {\bf u}\cdot\bec{\nabla} \biggr) {\bf u} = -\bec{\nabla}\biggl(p+{Q_p\over
8\pi}{\bf B}^2 \biggr)
+  ({\bf B}\cdot\bec{\nabla}){Q_s \over 4\pi}
{\bf B} + {\bf F}_{\nu} + {\bf F}_{ext} ,
\label{C2}\\
{\partial {\bf B}\over \partial t}&=&\bec{\nabla}{\bf \times}[{\bf u}
{\bf \times}{\bf B}  - \tilde \eta  \bec{\nabla}{\bf \times} {\bf B}] ,
\label{C3}\\
{\partial \rho\over \partial t} &+& \bec{\nabla}\cdot(\rho  {\bf
u})=0,
\label{C4}
\end{eqnarray}
where ${\bf u}$ is the mean velocity, $p = p_k+p_{_{T}}$, $\, p_k$ is the mean fluid pressure, $p_{_{T}}$ is the turbulent pressure, ${\bf F}_{ext}$ is the external force (for example, the gravitational force ${\bf F}_{ext} = \rho {\bf g})$, $\,  {\bf g}$ is the free-fall acceleration, ${\bf F}_\nu$ is the dissipation force owing to the molecular and turbulent viscosities, and $\tilde\eta$ is the magnetic diffusion (molecular plus turbulent).
The turbulent diamagnetic (or paramagnetic) drift velocity and the $\alpha$ effect are not included in Eq.~(\ref{C3}) because both $\alpha$  and the turbulent diamagnetic drift velocity are much smaller than the Alfv\'{e}n velocity for the range investigated. Next, we use the following identity:
\begin{eqnarray}
\rho \biggl({\partial \over \partial t} + {\bf u}\cdot\bec{\nabla} \biggr) u_i = {\partial (\rho \, u_i) \over \partial t} + \bec{\nabla}_j  (\rho \, u_i \, u_j),
\label{C9}
\end{eqnarray}
that allows us to rewrite Eq.~(\ref{C2}) in the following form:
\begin{eqnarray}
{\partial (\rho \, u_i) \over \partial t} &+& {\bf \nabla}_j  (\rho \, u_i \, u_j) = -{\bf \nabla}_i \biggl(p+{Q_p\over 8\pi}{\bf B}^2 \biggr)
+  ({\bf B}\cdot\bec{\nabla}){Q_s \over 4\pi}
{\bf B}_i + [{\bf F}_{\nu} + {\bf F}_{ext}]_i .
\label{C10}
\end{eqnarray}
In derivation of Eq.~(\ref{C9}) we used Eq.~(\ref{C4}).

Let us derive the budget equation for energy of the mean fields. We multiply Eq.~(\ref{C10}) by the velocity ${\bf u}$, Eq.~(\ref{C3}) by $(Q_s / 4 \pi) {\bf B}$, and add them. The result is given by:
\begin{eqnarray}
&& {\partial \over \partial t}\biggl({\rho {\bf u}^{2}\over 2}+
Q_s{{\bf B}^{2}\over 8\pi}\biggr)=-\bec{\nabla}\cdot \biggl[{\bf
u}\biggl({\rho {\bf u}^{2}\over 2}+p
+{Q_p- Q_s\over
8\pi}{\bf B}^{2}\biggr)+{Q_s\over 4\pi}{\bf
B}{\bf \times} ({\bf u}{\bf \times} {\bf B})\biggr] +
\nonumber\\
&& \quad +  ({\bf F}_{ext}\cdot{\bf u}) - D_{m} - D_T + \biggl(p
+{Q_p-Q_s\over 8\pi}{\bf B}^{2}\biggr) \, (\bec{\nabla}\cdot {\bf
u}) + {{\bf B}^{2}\over 8\pi} \, \biggl({\partial \over \partial t} + {\bf u}\cdot\bec{\nabla} \biggr) Q_s,
\nonumber\\
\label{C5}
\end{eqnarray}
where $D_T$ is the density of the power released  by the
turbulent viscosity and the turbulent magnetic diffusion,
$D_m$ is the density of the power released through
the molecular dissipation. Equation~(\ref{C5}) is the budget equation for the total energy of the large-scale flow and magnetic field  $\rho {\bf u}^{2} / 2 +Q_s{\bf B}^{2} / 8\pi$. The term $\propto (Q_p-Q_s) \, {\bf B}^{2} \, (\bec{\nabla}\cdot {\bf u})$ in RHS of Eq.~(\ref{C5}) is the work of turbulent magnetic stresses and pressure; the last term in RHS of this equation, which depends on time and spatial derivatives of $Q_s$, is the work of the pondermotor force. For instance, the term $\propto \partial Q_s/ \partial t$ is similar to that in the classical electrodynamics when the magnetic permeability depends on time. In the equation for the total turbulent energy these terms have opposite signs [see Eq.~(\ref{C16})], so that the equation for the total energy, including the energy of the mean fields, the total (kinetic and magnetic) turbulent energy, the potential energy and the internal energy, does not contains these terms. In derivation of Eq.~(\ref{C5}) we used the following identities:
\begin{eqnarray}
&& u_i \, \biggl[{\partial (\rho \, u_i) \over \partial t} + {\nabla}_j  (\rho \, u_i \, u_j) \biggr] = {\partial \over \partial t}\biggl({\rho {\bf u}^{2}\over 2}+Q_s{{\bf B}^{2}\over 8\pi}\biggr)
+ \bec{\nabla}\cdot \biggl({\bf u} \, {\rho {\bf u}^{2}\over 2} \biggr)
- {{\bf B}^{2}\over 8\pi} \, {\partial Q_s \over \partial t},
\label{C11}\\
&& {\bf u} \cdot \biggl[({\bf B}\cdot\bec{\nabla}){Q_s \over 4\pi}
{\bf B} -\bec{\nabla}\biggl({Q_p\over
8\pi}{\bf B}^2 \biggr) \biggr]
+ {Q_s \over 4\pi} {\bf B} \cdot [\bec{\nabla}{\bf \times}({\bf u} {\bf \times} {\bf B})]= -\bec{\nabla}\cdot \biggl[{\bf
u}\Big({\rho {\bf u}^{2}\over 2}
\nonumber\\
&& \quad +{Q_p- Q_s\over 8\pi}{\bf B}^{2}\Big) +{Q_s\over 4\pi}{\bf B}{\bf \times} ({\bf u}{\bf \times} {\bf B})\biggr]
+ {Q_p-Q_s\over 8\pi} \, {\bf B}^{2}\,  (\bec{\nabla}{\bf \cdot} {\bf u}) + {{\bf B}^{2}\over 8\pi} \, ({\bf u}{\bf \cdot}\bec{\nabla}) Q_s .
\label{C12}
\end{eqnarray}

Using Eq.~(\ref{C5}) we obtain budget equation for the total
energy (the sum of the energy of the mean fields, the potential energy and the internal energy). In particular, the equation for the entropy reads
\begin{eqnarray}
\rho T \biggl({\partial S \over \partial t} + ({\bf u}
{\bf \cdot}\bec{\nabla}) S \biggr) =I+ D_m + {\rho W_T\over t_T} -\bec{\nabla}{\bf \cdot}{\bf\Phi},
\label{C6}
\end{eqnarray}
where $S = \ln(p \rho^{-\gamma}) / \gamma$ is the entropy,
$\gamma$ is the ratio of the specific heats, $I$ is the external source of the thermal energy, $W_T$ is the density of the total (kinetic + magnetic) energy of the MHD turbulence, $t_T$ is the characteristic time of the dissipation of the turbulent energy into thermal energy (that is of the order of characteristic turbulent time), and ${\bf \Phi}$ is the total thermal flux. An expression for the internal energy $e$ is determined by the first principle of thermodynamics $de=TdS+(p_k/\rho ^{2})d\rho$. Using this equation and Eqs.~(\ref{C4}), (\ref{C6}) we get the following budget equation for the internal energy:
\begin{eqnarray}
{\partial \over \partial t}(\rho e) = I + {\rho W_T\over t_T}
+  D_m - p_k\bec{\nabla} {\bf \cdot} {\bf u} -\bec{\nabla}{\bf \cdot}[{\bf\Phi}+ {\bf u} \rho e].
\label{C15}
\end{eqnarray}
The budget equation for the turbulent energy $W_T$ reads:
\begin{eqnarray}
{\partial \over \partial t}(\rho W_T) &=& -\bec{\nabla} {\bf \cdot}
(\rho W_T {\bf u}) + D_T + I_T - {\rho W_T \over \tau} -
\biggl(p_{T} + {Q_p- Q_s\over 8\pi}{\bf B}^{2} \biggr) \, (\bec{\nabla} {\bf \cdot}{\bf u})
\nonumber\\
&& \quad -{{\bf B}^{2}\over 8\pi} \, \biggl({\partial \over \partial t} + {\bf u}{\bf \cdot}\bec{\nabla} \biggr) Q_s .
\label{C16}
\end{eqnarray}
Then the budget equation for the total energy after taking into account the MHD turbulence has the following form:
\begin{eqnarray}
{\partial \over \partial t} \biggl( {\rho {\bf u}^{2} \over 2} +
Q_s{{\bf B}^{2}\over 8\pi} + \rho \, e \biggr) = - \bec{\nabla}{\bf \cdot}{\bf q} + ({\bf F}_{ext}{\bf \cdot}{\bf u}) + I + I_T .
\label{C7}
\end{eqnarray}
The total energy flux ${\bf q}$ is given by
\begin{eqnarray}
{\bf q} = \rho {\bf u} \biggl ({{\bf u}^{2}\over 2} +
\epsilon + {p \over \rho } \biggr) + {Q_s\over 4\pi}{\bf B}{\bf \times}
({\bf u}{\bf \times} {\bf B})
+ {\bf u}{Q_p-Q_s\over 8\pi}{\bf B}^{2} +
{\bf \Phi} .
\label{C8}
\end{eqnarray}
If $Q_p \not= Q_s$, the MHD turbulence produces additional work.
It is converted into the energy of the large-scale flow
and magnetic field even in the absence of dissipation. Decrease of the elasticity of the large-scale magnetic field caused by the generation of the magnetic fluctuations in the MHD turbulence is essential for systems with the large magnetic Reynolds numbers. In this case the developed MHD turbulence can give rise to negative $Q_p$. Such values of $Q_p$ (i.e., negative) result in excitations of the large-scale MHD instabilities, which draw free energy from the turbulent motions.

\section{Governing equations in the TELF-Model and the results of numerical simulations}

We use the following compressible MHD conservation equations for mass, momentum, energy and magnetic flux:
\begin{eqnarray}
{\partial \rho \over \partial t} &+& \nabla_i (\rho \, u_i) = 0 ,
\label{A1} \\
{\partial (\rho \, u_i) \over \partial t} &+& \nabla_j \big(\rho \, u_i\, u_j + P_{ij} \big) = - \rho \, \nabla_i \, \varphi ,
\label{A2} \\
{\partial E \over \partial t} &+& \nabla_i \Big[E \, u_i + P_{ij} \, u_j -  (\kappa + \kappa_{_{T}}) \, \nabla_i T
+ {(\eta + \eta_{_{T}}) \over 4 \pi}  \, (\nabla_j B_i - \nabla_i B_j) \, B_j
\nonumber \\
&& + F_i^{\rm rad} \Big] = 0 ,
\label{A3} \\
{\partial B_i \over \partial t} &+& \nabla_j \Big[u_j \, B_i - u_i \, B_j - (\eta + \eta_{_{T}}) \,  (\nabla_j B_i
- \nabla_i B_j) \Big] =  0 ,
\label{A4}
\end{eqnarray}
where
\begin{eqnarray}
P_{ij} &=& \Big[p + {2 \mu \over 3} \, \nabla_k \, u_k + {1 - q_p\over 8 \pi} \, {\bf B}^2 \Big] \, \delta_{ij} + \rho \, \tau_{ij}- \mu \, (\nabla_j \,u_i
+ \nabla_i \,u_j) - {1 - q_s \over 4 \pi} \, B_i \, B_j
\nonumber \\
&& - \, {1 \over 8 \pi} \, \Big[2 q_e \, \hat z_i \, \hat z_j + q_a \, \cos \phi \, (\hat \beta_i \, \hat z_j + \hat \beta_j \, \hat z_i)\Big] {\bf B}^2   ,
\label{A5}\\
E &=& {1 \over 2} \, \rho \, {\bf u}^2 + \rho \, e + \rho \, \varphi
+ {1 - q_s\over 8 \pi} \, {\bf B}^2 ,
\label{A6}\\
\tau_{ij} &=& - 2 C_S \, \Delta^2\, |S| \, \Big[S_{ij} - {1 \over 3} \,
(\nabla_k \, u_k) \, \delta_{ij} \Big]
+ {2 \over 3} \, C_C \, \Delta^2\, |S|^2 \, \delta_{ij} ,
\label{A7}
\end{eqnarray}
$\eta$ is the magnetic diffusivity due to electrical conductivity of the fluid, $\eta_{_{T}} = u_{_{\Delta}} \, \Delta / 3$ is the turbulent magnetic diffusivity, $\kappa$ is the molecular heat diffusivity, $\kappa_{_{T}}$ is the turbulent heat diffusivity, $\mu$ is the dynamic viscosity, $u_{_{\Delta}}$ is the rms velocity,
$u_{_{\Delta}} = \Delta \, \big[S^2 + g \, |\nabla_z T| / T\big]^{1/2},$ at the scale $\Delta$ (which is the computational grid step size),
$\hat \beta_i = B_i/B$, $\, \hat z_i$ is the vertical unit vector directed opposite to the gravity field, and $\phi$ is the angle
between the vertical unit vector $\hat{\bf z}$ and the magnetic field ${\bf B}$.

The formulae for the nonlinear functions, $q_{\rm p}(\beta)$, $\, q_{\rm s}(\beta)$, $\, q_{\rm e}(\beta)$, and $q_{\rm a}(\beta)$, are given in Rogachevskii \& Kleeorin (2007), where $\beta \equiv B/B_{\rm eq}$ is the non-dimensional magnetic field, $B_{\rm eq}=\sqrt{\rho} \, u_{_{\Delta}}$ is the equipartition field strength based on rms velocity $u_{_{\Delta}}$, and the magnetic Reynolds number at this scale ${\rm Rm}_{_{\Delta}} = u_{_{\Delta}} \, \Delta / \eta$. In LES discussed in the present study we used the following fitting formulae for these nonlinear functions based on parameterizations both of analytic formulae by Rogachevskii \& Kleeorin (2007) and the results of the DNS (Brandenburg {\it et al.} 2010a,b; K\"{a}pyl\"{a} {\it et al.} 2010):
\begin{eqnarray}
q_{\rm p} &=& C_1 \, \left(1 - {2\over\pi} \arctan{C_2 \, B^2\over B_{\rm eq}^2} \right),
\quad q_{\rm s} = q_{\rm p} / 2, \quad q_{\rm e} = q_{\rm a} = 0,
\label{A9}
\end{eqnarray}
where the parameters can be chosen as $C_1 = 50$ and $C_2 = 25$.

The numerical simulation results are obtained for the computational
domain of $6.4 \times 6.4 \times 5.5$ Mm with the grid sizes $50 \times 50 \times 43$ km ($256^3$ mesh points). For details see the numerical set-up of Kitiashvili {\it et al.} (2010). The domain includes a top,
5 Mm - deep layer of the convective zone, and, 0.5 Mm - deep layer of the low atmosphere. The lateral boundary conditions are periodic, and the top and bottom boundaries are closed to mass, momentum, and energy
transfer (apart from radiative losses). There is no net
magnetic field lost or gained through the boundaries, i.e., the
integral over volume for each component of the field is constant.
The initial uniform vertical and horizontal magnetic fields of various magnitudes (1, 10, and 100 G) were superimposed on the fully developed granular convection. The computation runs were up to 3 h of solar time.

\begin{figure}
\begin{center}
\includegraphics[scale=0.8]{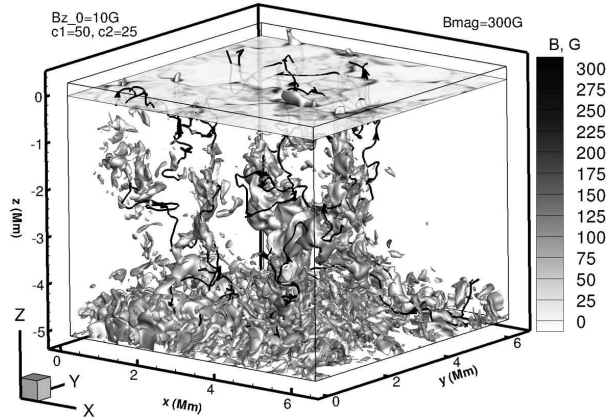}
\end{center}
\caption{\label{10v} Instantaneous distribution of a magnetic field,  taking into account the effect of turbulent convection on large-scale Lorentz force obtained in LES with the TELF-Model for the case of vertical initial magnetic field 10 G.}
\end{figure}

\begin{figure}
\begin{center}
\includegraphics[scale=0.8]{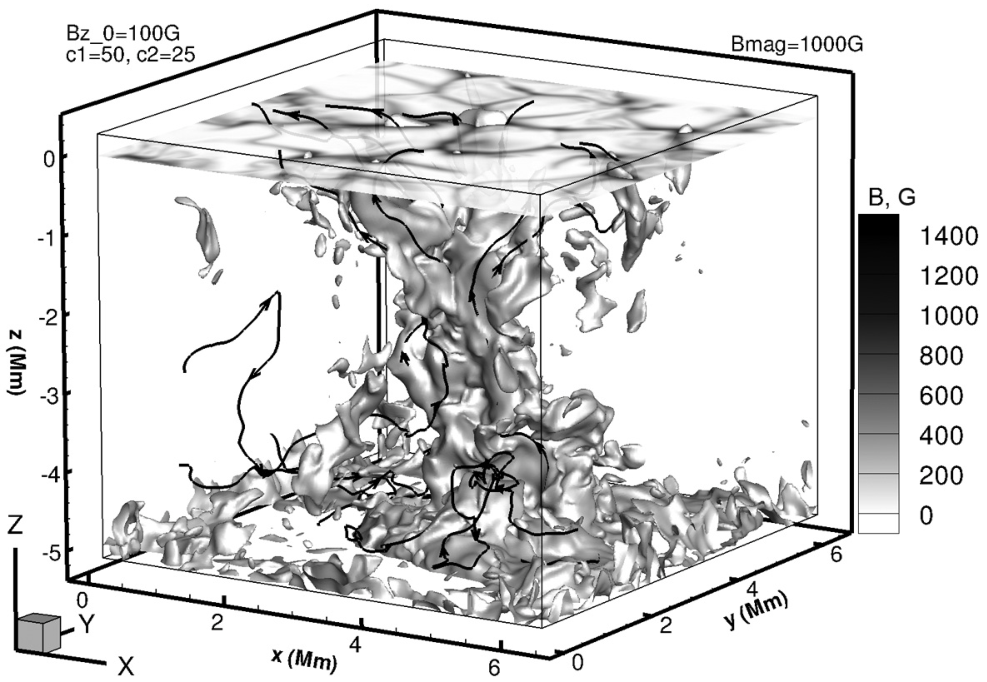}
\end{center}
\caption{\label{100v} Instantaneous distribution of a magnetic field obtained in LES with the TELF-Model for the case of vertical initial magnetic field 100 G.}
\end{figure}

\begin{figure}
\begin{center}
\includegraphics[scale=0.8]{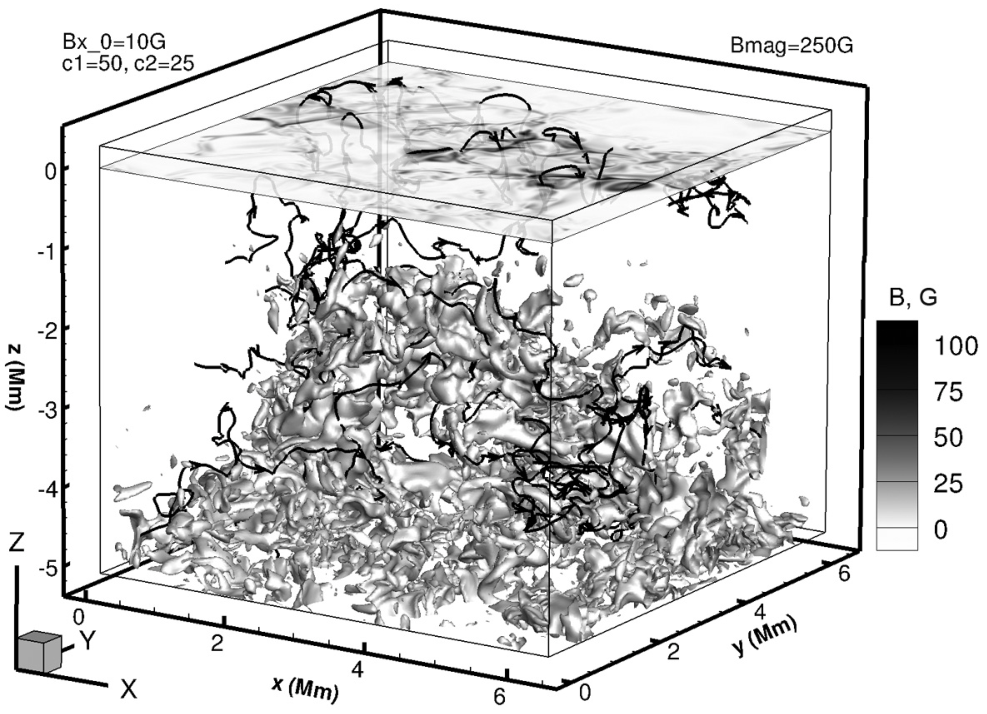}
\end{center}
\caption{\label{10h} Instantaneous distribution of a magnetic field obtained in LES with the TELF-Model  for the case of horizontal initial magnetic field 10 G.}
\end{figure}

\begin{figure}
\begin{center}
\includegraphics[scale=0.8]{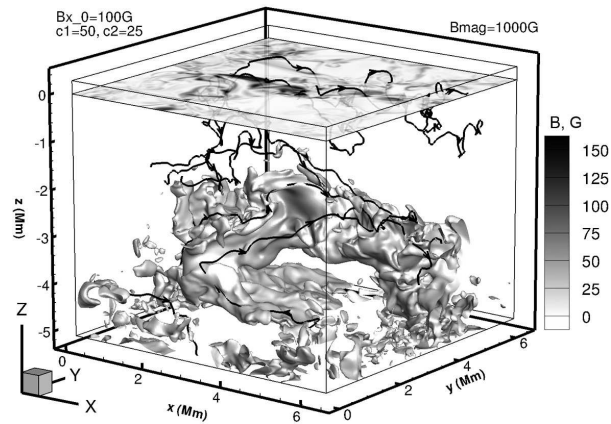}
\end{center}
\caption{\label{100h} Instantaneous distribution of a magnetic field obtained in LES with the TELF-Model  for the case of horizontal initial magnetic field 100 G.}
\end{figure}

The results of the numerical simulations are given below,
whereby we show the instantaneous distributions of magnetic field obtained in LES, with the TELF-model, that takes into account the effect of subgrid turbulent convection on large-scale Lorentz force. We considered two cases of the initial vertical (see Figs.~\ref{10v}-\ref{100v}) and horizontal (see Figs.~\ref{10h}-\ref{100h}) uniform magnetic field. These figures show that the magnetic flux tubes formation in a fully developed turbulent convection depends on the magnitude of the initial magnetic field (e.g., when the initial uniform magnetic field is about 100 G, the magnetic flux tubes are formed; see Figs.~\ref{100v} and~\ref{100h}). This implies that the process of the magnetic flux tubes formation is a threshold effect.

\begin{figure}
\begin{center}
\includegraphics[scale=0.35]{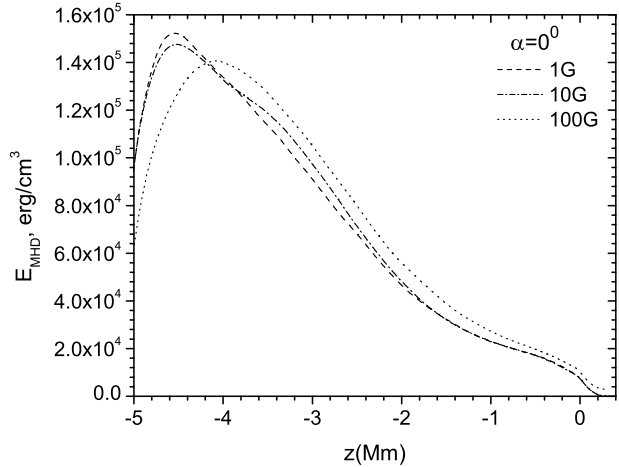}
\end{center}
\caption{\label{MHD_energy} Vertical distribution of the turbulent magnetohydrodynamic energy $\rho \langle {\bf u}^{2} \rangle/ 2 + \langle {\bf b}^{2} \rangle/ 8 \pi$  for the case of vertical initial magnetic fields of different magnitudes.}
\end{figure}

\begin{figure}
\begin{center}
\includegraphics[scale=0.35]{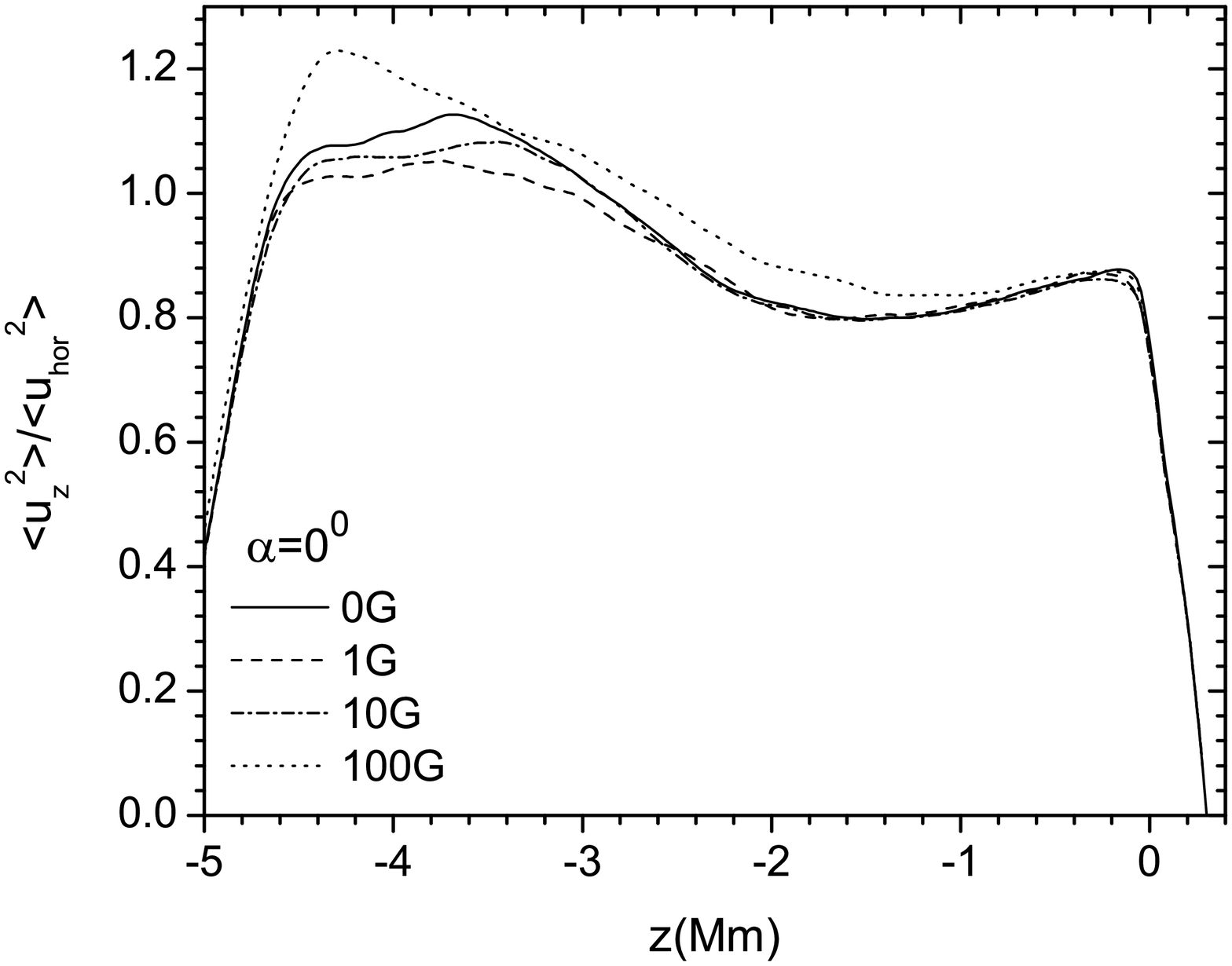}
\end{center}
\caption{\label{V_anisotropy} Vertical distribution of the anisotropy parameter of a turbulent velocity field $\langle {\bf u}_z^{2} \rangle / \langle {\bf u}_{\rm hor}^{2} \rangle$  for the case of vertical initial magnetic fields of different magnitudes.}
\end{figure}

\begin{figure}
\begin{center}
\includegraphics[scale=0.35]{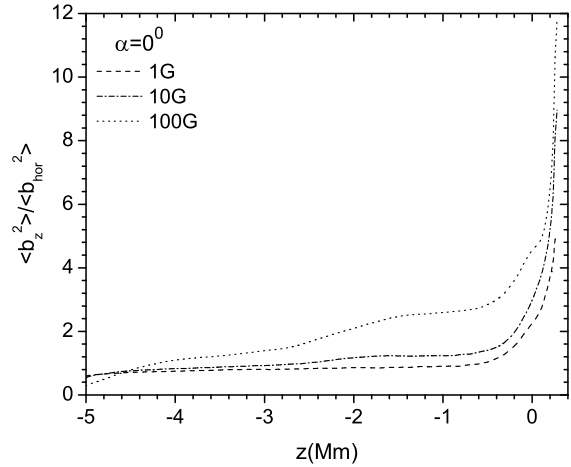}
\end{center}
\caption{\label{b_anisotropy} Vertical distribution of the anisotropy parameter of a turbulent magnetic field $\langle {\bf b}_z^{2} \rangle / \langle {\bf b}_{\rm hor}^{2} \rangle$  for the case of vertical initial magnetic fields of different magnitudes.}
\end{figure}

In this study we determined the different statistical characteristic of the turbulent convection and formed magnetic field, e.g., we find the vertical profiles of the velocity and magnetic fluctuations, MHD energy, anisotropy of turbulent magneto-convection, and kinetic, current and cross helicities. In particular, in Figs.~\ref{MHD_energy}-\ref{b_anisotropy} we show the vertical distributions of the turbulent magnetohydrodynamic energy $\rho \langle {\bf u}^{2} \rangle/ 2 + \langle {\bf b}^{2} \rangle/ 8 \pi$ and the anisotropy parameters of velocity fluctuations, $\langle {\bf u}_z^{2} \rangle / \langle {\bf u}_{\rm hor}^{2} \rangle$, and magnetic  fluctuations, $\langle {\bf b}_z^{2} \rangle / \langle {\bf b}_{\rm hor}^{2} \rangle$. Owing to a strong fluid density contrast, the main turbulent energy is located at the low part of the convective layer. On the other hand, the anisotropy parameters of velocity and magnetic fluctuations are distributed nearly uniformly in the convective layer.

\section{Conclusions}

In this study we derived the energy budget equations that include the effects of the small-scale (e.g., subgrid-scale) turbulence on the Lorentz-force for modelling of formation of the magnetic self-organized structures in the realistic solar conditions. We implemented the new subgrid-scale turbulence model (TELF-Model) in a three-dimensional nonlinear MHD LES code and performed LES simulations with the TELF-Model with imposed initial vertical and horizontal uniform magnetic fields.
We have shown that the magnetic flux tubes formation is driven when the initial mean magnetic field is larger than a threshold value, in  agreement with the theoretical studies by Rogachevskii \& Kleeorin (2007). They demonstrated that the magnetic flux tubes formation is caused by the large-scale MHD instability and that a threshold in the initial magnetic field is required for the excitation of this instability. We also determined the vertical profiles of the velocity and magnetic fluctuations, MHD energy, anisotropy of turbulent magneto-convection, kinetic and current and cross helicities.

In the future study of magnetic flux concentrations one need to determine in LES the turbulent Reynolds and Maxwell stress tensors in turbulent convection as a function of the initial magnetic field imposed on a turbulent convection. From the latter, one will be able to know, from LES, the turbulence contributions to the mean Lorentz force and compare them with theoretical predictions and DNS results.
It is also important to study the influence of different effects (e.g., inclination of the initial magnetic field, differential rotation, meridional circulations, etc.) on the formation and decay of the inhomogeneous magnetic structures. It might be worthwhile to perform LES with larger scale separation, and over a broader range of values of the parameters. In this way it should be possible to optimize our choice of parameters in order to see where the results of mean-field, DNS, and LES can be brought closer together.

This study might be important for better understanding of the fundamental  mechanisms governing the formation of sunspots and active regions, and it is expected to have an impact in the modelling and predictions of solar magnetic activity.

We gratefully acknowledge support from Center for Turbulence Research.


\begin{thebibliography}{15}
\expandafter\ifx\csname natexlab\endcsname\relax\def\natexlab#1{#1}\fi

\bibitem[Brandenburg et al. (2010a)]{BKR2010a} {\sc Brandenburg A., Kemel K., Kleeorin N. \& Rogachevskii I.} 2010a Effect of stratified turbulence on magnetic flux concentrations. {\em Astrophys. J.}, submitted; arXiv:1005.5700.

\bibitem[Brandenburg et al. (2010b)]{BKR2010b} {\sc Brandenburg A., Kleeorin N. \& Rogachevskii I.} 2010b Large-scale magnetic flux concentrations from turbulent stresses. {\em Astron. Nachr.\/} {\bf 331}, 5--13.


\bibitem[Germano et al. (1991)]{Germano1991} {\sc Germano M., Piomelli U., Moin P. \& Cabot W.~H.} 1991 A dynamic subgrid-scale eddy viscosity model. {\em Phys. Fluids} {\bf 3}, 1760--1765.

\bibitem[Jacoutot et al. (2008a)]{Jacoutot2008a} {\sc Jacoutot L., Kosovichev A.~G., Wray A. \& Mansour N.~N.} 2008a Numerical simulation of excitation of
    solar oscillation modes for different turbulent models. {\em Astrophys. J.} {\bf 682}, 1386--1391.

\bibitem[Jacoutot et al. (2008b)]{Jacoutot2008b} {\sc Jacoutot L., Kosovichev A.~G., Wray A. \& Mansour N.~N.} 2008b Realistic numerical simulations of solar
    convection and oscillations in magnetic regions. {\em Astrophys. J.} {\bf 684}, L51--L54.


\bibitem[K\"{a}pyl\"{a} et al. (2010)]{KBKR2010} {\sc K\"{a}pyl\"{a} P.~J., Brandenburg A., Kleeorin N., Korpi M.~J. \& Rogachevskii I.} 2010c Effect of turbulent convection on mean Lorentz force and magnetic flux concentrations. {\em Monthly Not. Roy. Astron. Soc.}, to be submitted.

\bibitem[Kitiashvilli et al. (2009)]{Kitiashvilli2009} {\sc Kitiashvilli I., Kosovichev A.~G., Wray A. \& Mansour N.~N.} 2009 Traveling waves of magnetoconvection
    and the origin of the Evershed effect in sunspots. {\em Astrophys. J.} {\bf 700}, L178--L181.

\bibitem[Kitiashvilli et al. (2010)]{Kitiashvilli2010} {\sc Kitiashvilli I., Kosovichev A.~G., Wray A. \& Mansour N.~N.} 2010 Mechanism of spontaneous formation of stable magnetic structures on the Sun. {\em Astrophys. J.} {\bf 719}, 307--312.


\bibitem[Kleeorin et al. (1996)]{KMR1996} {\sc Kleeorin N., Mond M. \& Rogachevskii I.} 1996 Magnetohydrodynamic turbulence in the solar convective zone as a source of oscillations and sunspots formation. {\em Astron. Astrophys.} {\bf 307}, 293---309.

\bibitem[Kleeorin \& Rogachevskii (1994)]{KR1994} {\sc Kleeorin N. \& Rogachevskii I.} 1994 Effective Amp\`{e}re force in developed magnetohydrodynamic turbulence.  {\em Phys. Rev. E} {\bf 50}, 2716--2730.

\bibitem[Moin et al. (1991)]{Moin1991} {\sc Moin P., Squires K., Cabot W. \& Lee S.} 1991  A dynamic subgrid-scale model for compressible turbulence and scalar transport. {\em Phys. Fluids A} {\bf 3}, 2746--2757.

\bibitem[Rempel et al. (2009)]{Rempel2009} {\sc Rempel, M., Sch\"{u}ssler, M., Cameron, R.H. \& Kn\"{o}lker, M.} 2009, Penumbral structure and outflows in simulated sunspots. {\em Science} {\bf 325}, 171--174.

\bibitem[Rogachevskii \& Kleeorin (2007)]{RK2007} {\sc Rogachevskii I. \& Kleeorin N.} 2007 Magnetic fluctuations and formation of large-scale inhomogeneous magnetic structures in a turbulent convection. {\em Phys. Rev. E} {\bf 76}, 056307.

\bibitem[Smagorinsky (1963)]{Smagorinsky1963} {\sc Smagorinsky J.} 1963 General circulation experiments with the primitive equations. {\em Monthly Weather Rev.} {\bf 93}, 99--164.

\bibitem[Stein et al. (2010)]{Stein2010} {\sc Stein, R. F., Lagerfj\"{a}rd, A., Nordlund, {\AA}.\& Georgobiani, D.} 2010, Solar flux emergence simulations. {\em Sol. Phys.} {\bf 34}, No. 2.

\bibitem[Theobald et al. (1994)]{Theobald1994} {\sc Theobald M. L., Fox P. A. \& Sofia S.} 1994 A subgrid-scale resistivity for magnetohydrodynamics. {\em Phys. Plasmas} {\bf 1}, 3016--3032.

\end{thebibliography}
\end{document}